# Implementation of Multilayer Perceptron Network with Highly Uniform Passive Memristive Crossbar Circuits

F. Merrikh Bayat[1], M. Prezioso[1], B. Chakrabarti[1], I. Kataeva[2], and D. Strukov[1]

[1] Electrical and Computer Engineering Department, University of California Santa Barbara, Santa Barbara, CA, 93117, USA

[2] Research Laboratories, DENSO CORP., 500-1 Minamiyama, Komenoki-cho, Nisshin, Japan 470-0111

**The progress in the field of neural computation hinges on the use of hardware more efficient than the conventional microprocessors. Recent works have shown that mixed-signal integrated memristive circuits, especially their passive ("0T1R") variety, may increase the neuromorphic network performance dramatically, leaving far behind their digital counterparts. The major obstacle, however, is relatively immature memristor technology so that only limited functionality has been demonstrated to date. Here we experimentally demonstrate operation of one-hidden layer perceptron classifier entirely in the mixed-signal integrated hardware, comprised of two passive 20×20 metal-oxide memristive crossbar arrays, board-integrated with discrete CMOS components. The demonstrated multilayer perceptron network, whose complexity is almost 10× higher as compared to previously reported functional neuromorphic classifiers based on passive memristive circuits, achieves classification fidelity within 3% of that obtained in simulations, when using ex-situ training approach. The successful demonstration was facilitated by improvements in fabrication technology of memristors, specifically by lowering variations in their *I-V* characteristics.**

Started more than half a century ago, the field of neural computation has known its ups and downs, but since 2012, it exhibits an unprecedented boom triggered by the dramatic breakthrough in the development of deep convolutional neuromorphic networks[1,2]. The breakthrough[3] was enabled not by any significant algorithm advance, but rather by the use of high performance graphics processors[4], and the further progress is being fueled now by the development of even more powerful graphics processors and custom integrated circuits[5-7]. Nevertheless, the energy efficiency of these implementations of convolutional networks (and other neuromorphic systems[8-11]) remains well below that of their biological prototypes[12,13], even when the most advanced CMOS technology is used. The main reason for this efficiency gap is that the use of digital operations for mimicking biological neural networks, with their high redundancy and intrinsic noise, is inherently unnatural. On the other hand, recent works have shown[11-16] that analog and mixed-signal integrated circuits, especially using nanoscale devices, may increase the neuromorphic network performance dramatically, leaving far behind both their digital counterparts and biological prototypes and approaching the energy efficiency of the brain. The background for these advantages is that in such circuits the key operation performed by any neuromorphic network, the vector-by-matrix multiplication, is implemented on the physical level by utilization of the fundamental Ohm and Kirchhoff laws. The key component of this circuit is a nanodevice with adjustable conductance $G$ - essentially an analog nonvolatile memory cell - used at each crosspoint of a crossbar array, and mimicking the biological synapse.

Though potential advantages of specialized hardware for neuromorphic computing had been recognized several decades ago[17,18], up until recently, adjustable conductance devices were mostly implemented using the standard CMOS technology[13]. This approach was used to implement several sophisticated, efficient systems – see, e.g. Refs. [14, 15]. However, these devices have relatively large areas leading to higher interconnect capacitance and hence larger time delays. Fortunately, in the last decade, another revolution has taken place in the field of nanoelectronic memory devices. Various types of emerging nonvolatile memories are now being actively investigated for their use in fast and energy-efficient neuromorphic networks[19-41]. Of particular importance, is the development of the technology for programmable, nonvolatile two-terminal devices called ReRAM or "memristors" [42,43]. The low-voltage conductance $G$ of these devices may be continuously adjusted by the application of short voltage pulses of higher amplitude (>1 V) [27,42]. These devices were used to demonstrate first simple neuromorphic network providing pattern classification [21,26,27,28,30,32,40]. The memristors can have a very low chip footprint, which is determined only by the overlap area of the metallic electrodes, and may be scaled down below 10 nm without sacrificing their endurance, retention, and tuning accuracy, with some of the properties (such as the ON/OFF conductance ratio) being actually improved [44].

The main result of this paper is the experimental demonstration of a fully functional, board-integrated, mixed-signal neuromorphic network based on integrated metal-oxide memristive devices. The demonstrated network is comprised of almost an order of magnitude higher number of devices as compared to the previously reported neuromorphic classifiers based on passive crossbar circuits.[26,27] The inference, the most common operation in applications of deep learning, is performed directly in a hardware, which is different from many previous works that relied on post-processing the experimental data with external computer to emulate the functionality of the whole system.[25-27,34,39,40.] Our particular focus is on passive ("0T1R") memristive crossbar circuits, which are naturally suitable for three-dimensional integration[45-47]. Due to their extremely high effective integration density, such circuits may be instrumental for keeping all the synaptic weights of a large-scale artificial neural networks locally, thus cutting dramatically the energy and latency overheads of the off-chip communications.

## Integrated Memristors

The passive 20×20 crossbar arrays with $Pt/Al_2O_3/TiO_{2-x}/Ti/Pt$ memristor at each crosspoint were fabricated using a technique similar to that reported in Refs. 26, 27 (Fig. 1). Specifically, the bilayer binary oxide stack was deposited using low temperature reactive sputtering method. The crossbar electrodes were evaporated using oblique angle physical vapor deposition (PVD) and patterned by lift-off technique using lithographical masks with 200-nm lines separated by 400-nm gaps. Each crossbar electrode is contacted to a thicker (Ni/Cr/Au 400 nm) metal line / bonding pad, which are formed at the last step of the fabrication process. As evident from Figure 1a, b, due to the utilized undercut in the photoresist layer and tilted PVD sputtering in the lift-off process, the metal electrodes have roughly triangular shape with ~ 250 nm width. Such shape of the bottom electrodes ensured better step coverage for the following processing layers and, in particular, helped to reduce the top electrode resistance. The externally measured (pad-to-pad) crossbar line resistance for the bonded chip is around 800 Ω. It is similar to that of smaller crossbar circuit reported in Refs. 26, 27 due to the dominant contribution of the contact between crossbar electrode and thicker bonding lines.

Majority of the devices required an electroforming step which consisted of one-time application of a high current ramp bias. The devices were formed one at a time, and to speed up the whole process, an automated setup has been developed (Fig. S1a). Such setup was used for early screening of defective samples, and has allowed a successful forming and testing of numerous crossbar arrays (Fig. 2).

Memristor *I-V* characteristics are nonlinear (Fig. 1c) due to the alumina barrier between the bottom electrode and the switching layer. *I-V*'s nonlinearity provides selector functionality, which is essential for the conductance tuning in the crossbar circuit. In particular, it limits leakage currents and reduces disturbance of half-selected devices – see Sect. 3 of the Supplementary Information for more discussion of this point.

Most importantly, memristive devices in the fabricated 20×20 crossbar circuits have uniform characteristics with gradual "analog" switching. Specifically, the distributions of the effective set and reset voltages are sufficiently narrow (Fig. 2) to allow precise tuning of devices' conductances to the desired values in the whole array (Fig. 3). To the best of our knowledge, this is the first report of such a precise adjustment on this integration scale. For example, an analog tuning was essential for other demonstrations based on passive memristive circuits, though was performed with much cruder precision [19,39]. A comparable tuning accuracy was demonstrated in Ref. 40, though for less dense but much more robust to variations "1T1R" structures, in which each memory cell is coupled with a dedicated transistor.

## Multilayer Perceptron Implementation

Two 20×20 crossbar circuits were packaged and integrated with discrete CMOS components on two printed circuit boards (Fig. S2b) to implement the multilayer perceptron (MLP) (Fig. 4). The MLP network features 16 inputs, 10 hidden-layer neurons, and 4-outputs, which is sufficient to perform classification of 4×4-pixel black-and-white patterns (Fig. 4d) into 4 classes. With account of bias inputs, the implemented neural network has 170 and 44 synaptic weights in the first and second layers, respectively.

The integrated memristors implement synaptic weights, while discrete CMOS circuitry implements switching matrix and neurons. Each synaptic weight is implemented with a pair of memristors, so that 17×20 and 11×8 contiguous subarrays were involved in the experiment (Fig. 4a), i.e. almost all of the available memristors in the first crossbar and about a quarter of the devices in the second one. The switching matrix was implemented with analog discrete component multiplexers and designed to operate in two different modes. The first one is utilized for on-board forming of memristors as well as their conductance tuning during weight import. In this operation mode, the switching matrix allows the access to any selected row and column and, simultaneously, the application of a common voltage to all remaining (half-selected) crossbar lines, including an option of floating them. The voltages are generated by an external parameter analyzer. In the second, inference mode the switching matrix connects the crossbar circuits to the neurons as shown in Fig. 4a and enables the application of ±0.2 V inputs, corresponding to white and black pixels of the input patterns. Concurrently, the measurement of output voltages of the perceptron network is carried out. The whole setup is controlled by a general-purpose computer (Fig. S2c).

The neuron circuitry is comprised of three distinct stages (Fig. S2a). The first stage consists of inverting operational amplifier, which maintains a virtual ground on the crossbar row electrodes. Its voltage output is a weighted sum between the input voltages, applied to crossbar columns (Fig. 4a), and the conductances of the corresponding crosspoint devices. The second stage op-amp computes the difference between two weighted sums calculated for the adjacent rows of the crossbar. The operational amplifier’s output in this stage is allowed to saturate for large input currents, thus effectively implementing *tanh*-like activation function. In the third and final stage of the neuron circuit, the output voltage is scaled down to be within -0.2 V to +0.2 V range before applying it to the next layer. The voltage scaling is only implemented for the hidden layer neurons to ensure negligible disturbance of the state of memristors in the second crossbar array.

With such implementation, perceptron operation for the first and second layers is described by the following equations:

$$V_j^{\mathrm{H}} \approx 0.2\tanh\left[10^6\left(I_j^+ - I_j^-\right)\right], \qquad I_j^{\pm} = \sum_{i=1}^{17} V_i^{\mathrm{in}} G_{ij}^{(1)\pm} \tag{1}$$

$$V_k^{\text{out}} \approx 10^6(I_k^+ - I_k^-), \qquad I_k^\pm = \sum_{j=1}^{11} V_j^{\text{H}} G_{jk}^{(2)\pm} \tag{2}$$

Here $V^{\text{in}}$, $V^{\text{H}}$, $V^{\text{out}}$ are, respectively, perceptron input, hidden layer output, and perceptron output voltages. $G^{(1)\pm}$ and $G^{(2)\pm}$ are the device conductances in the first and second crossbar circuits, with +/- superscripts denoting a specific device of a differential pair, while $I^\pm$ are the currents flowing into the corresponding neurons. $1 \leq j \leq 10$ and $1 \leq k \leq 4$ are hidden and output neuron indexes, while $1 \leq i \leq 16$ is the pixel index of an input pattern. The additional bias inputs are $V_{17}^{\text{in}} = V_{11}^{\text{H}} \equiv +0.2$ V.

## Pattern Classification

The multilayer perceptron was trained ex-situ by first finding the synaptic weights in the software-implemented network, and then importing the weights into the hardware. (Some of the preliminary results for in-situ training are discussed in Sect. 2 of the Supplementary Information.) In particular, the software-based perceptron was trained with conventional backpropagation algorithm, using four sets of patterns representing four classes of letters (Fig. 4d). In the software network, the neuron activation function was approximated with tangent hyperbolic with a slope specific to the hardware implementation. We assumed a linear *I*-*V* characteristics for the memristors, which is a good approximation for the considered range of voltages used for inference operation (Fig. 1c). During the training the weights were clipped within [10 μS, 100 μS] conductance range, which is an optimal range for the considered memristors.

In addition, two different approaches for modeling weights were considered in the software network. In the simplest "hardware-oblivious" approach, all memristors were assumed to be perfectly functional, while in a more advanced "hardware-aware" approach, the software model utilized additional information about the defective memristors. These were the devices whose conductances were experimentally found to be stuck at some values, and hence could not be changed during tuning.

The calculated synaptic weights were imported into the hardware by tuning memristors' conductances to the desired values using an automated write-and-verify algorithm.[48] The stuck devices were excluded from tuning for the hardware-aware training approach. To speed up weight import, the maximum tuning error was set to 30% of the target conductance (Fig. 5a, b), which is adequate import precision for the considered benchmark according to the simulation results (Fig. S5). After weight import had been completed, the inference was performed by applying ±0.2V inputs specific to the pattern pixels and measuring four analog voltage outputs. Fig. 5c shows typical transient response. Though the developed system was not optimized for speed, the experimentally measured classification rate was quite high – about 300,000 patterns per second and was mainly limited by the chip-to-chip propagation delay of analog signals on the printed circuit board.

Figure 5d,e shows classification results for the considered benchmark using the two different approaches. (In both software simulations and hardware experiments, the winning class was determined by the neuron with maximum output voltage.) The generalization functionality was tested on a 640 noisy test patterns (Fig. S4), obtained by flipping one of the pixels in the training images (Fig. 4d). The experimentally measured fidelity on a training and test set patterns for the hardware-oblivious approach were respectively 95% and 79.06% (Fig. 5d), as compared to 100% and 82.34% achieved in the software (Fig. S5). As expected, the experimental results were much better for hardware-aware approach, i.e. 100% for the training patterns and 81.4% for the test ones (Fig. 5e).

It should be noted that the achieved classification fidelity on test patterns is far from ideal 100% value due to rather challenging benchmark. In our demonstration, the input images are small and addition of noise, by flipping one pixel, resulted in many test patterns being very similar to each other. In fact, many of them are very difficult to classify even for a human, especially distinguishing between test patterns "V" and "X".

## Discussion and Summary

We believe that the presented work is an important milestone towards implementation of extremely energy efficient and fast mixed-signal neuromorphic hardware. Though demonstrated network has rather low complexity to be useful for practical applications, it has all major features of more practical large-scale deep learning hardware – a nonlinear neuromorphic circuit based on metal-oxide memristive synapses integrated with silicon neurons. The successful board-level demonstration was mainly possible due to the advances in memristive circuit fabrication technology, in particular much improved uniformity and reliability of memristors.

Perhaps the only practically useful way to scale up the neuromorphic network complexity further is via monolithical integration of memristors with CMOS circuits. Such work has already been started by several groups[19,30], including ours[47]. We envision that the most promising implementations will be based on passive memristor technology, i.e. similar to the one demonstrated in this paper, because it is suitable for monolithical back-end-of-line integration of multiple crossbar layers[46]. The three dimensional nature of such circuits[49] will enable neuromorphic networks with extremely high synaptic density, e.g. potentially reaching $10^{13}$ synapses in one square centimeter for 100-layer 10-nm memristive crossbar circuits, which is only hundred times less compared to the total number of synapses in a human brain. (Reaching such extremely high integration density of synapses would also require substantially increasing crossbar dimensions - see discussion of this point in Section 3 of Supplementary Information.) Storing all network weights locally would eliminate overhead of the off-chip communication and lead to unprecedented system-level energy efficiency and speed for large-scale networks. For example, the crude estimates showed that energy-delay product for the inference operation of a large-scale deep learning neural networks implemented with mixed-signal circuits based on the 200-nm memristor technology similar to the one discussed in this paper could be six orders of magnitude smaller as compared to that of the advanced digital circuits, while more than eight orders of magnitude smaller when utilizing three-dimensional 10-nm memristor circuits[50].

## Acknowledgements

This work was supported by DARPA under contract HR0011-13-C-0051UPSIDE via BAE Systems, Inc., by NSF grant CCF-1528205, and by the DENSO CORP., Japan. Useful discussions with G. Adam, B. Hoskins, X. Guo, and K. K. Likharev are gratefully acknowledged.

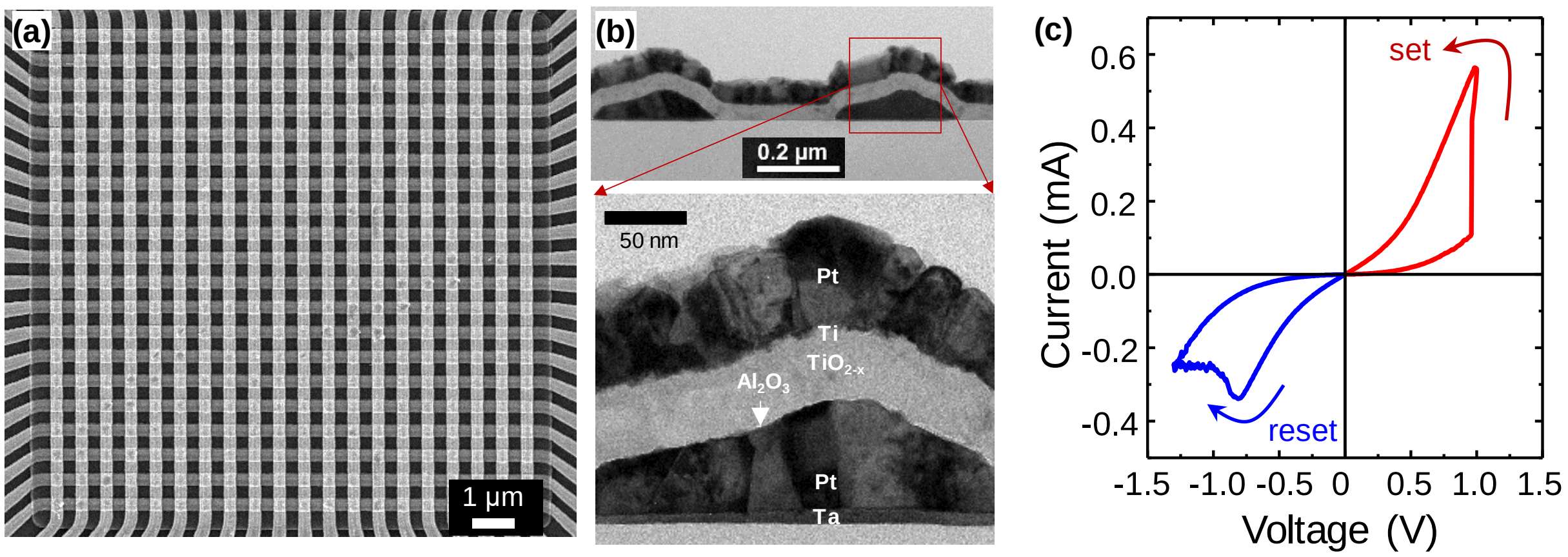

**Figure 1.** Passive 20×20 crossbar circuit with integrated Pt/$Al_2O_3$/$TiO_{2-x}$/Ti/Pt memristors: (a) A top-view SEM and (b) cross-section TEM images; (c) A typical *I-V* switching curve.

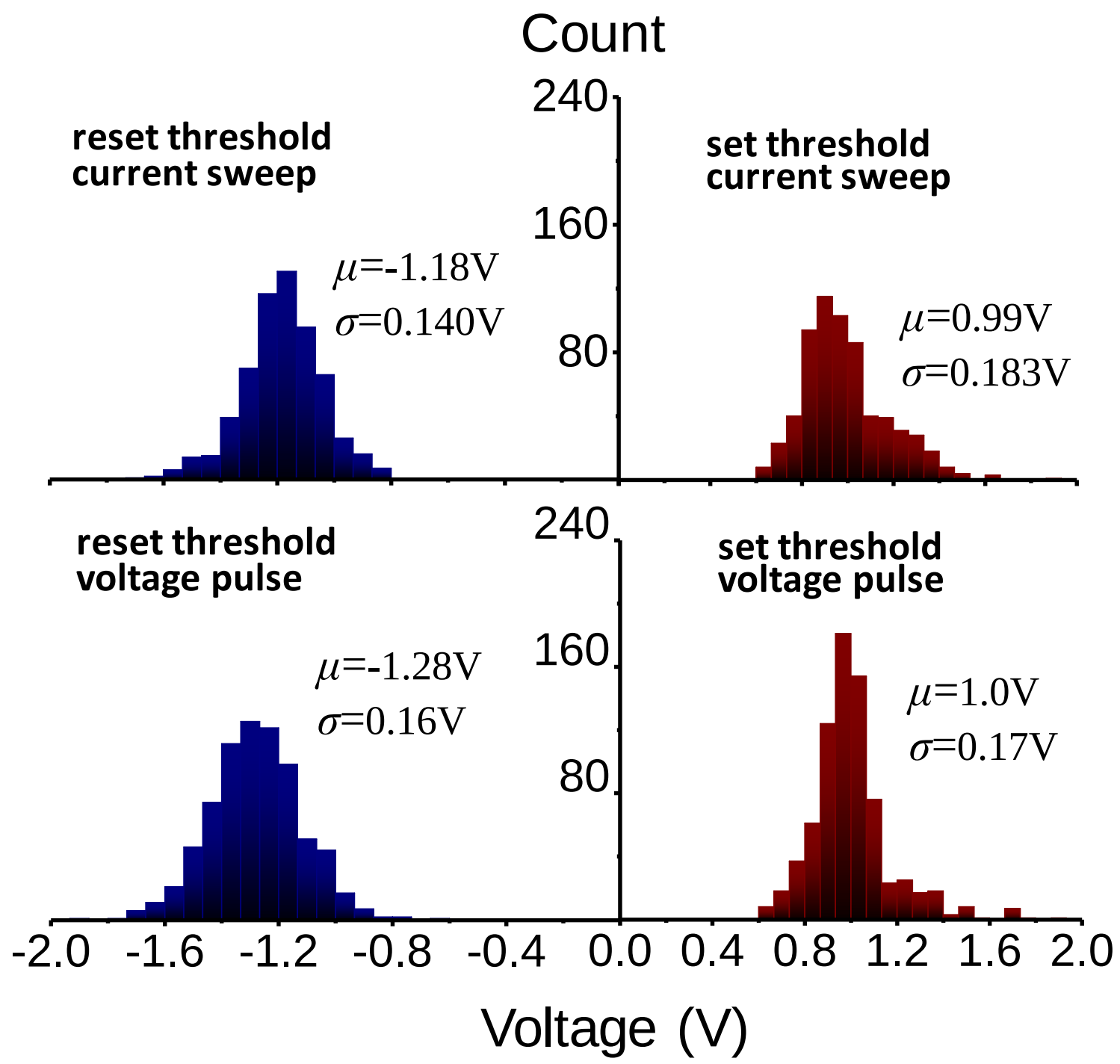

**Figure 2.** Set and reset threshold statistics for seven 20×20- device crossbar arrays at memristor switching with current and voltage pulses. The set / reset thresholds are defined as the smallest voltages at which the device resistance is increased / decreased by more than 5% at the application of a voltage or current pulse of the corresponding polarity. The legends show the corresponding averages and standard deviations for the switching threshold distributions. Note that the variations are naturally better when only considering devices within a single crossbar circuit, and in addition, excluding memristors at the edges of the circuit, which typically contribute to the long tails of the histograms. For example, $\mu$ = 1.0 V and σ = 0.13 V for voltage controlled set, while it is $\mu$ = -1.2 V and σ = 0.15 V for reset for one of the crossbars used in the experiment.

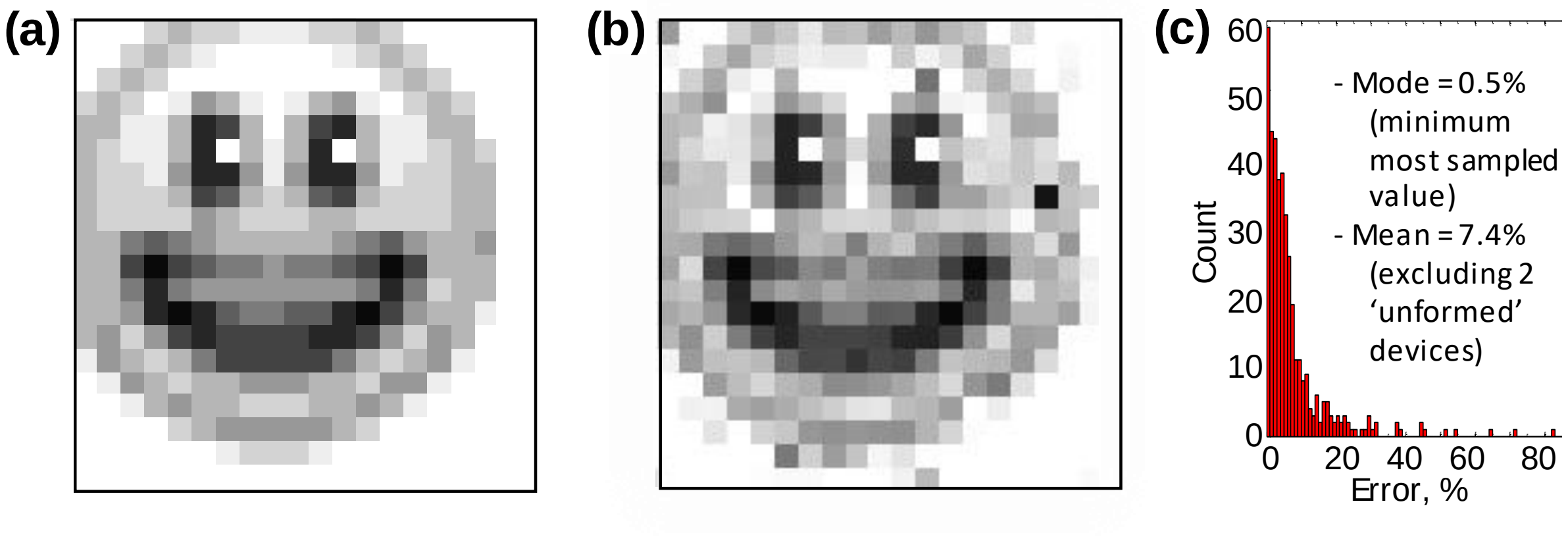

**Figure 3.** High precision tuning in 20×20 memristive crossbar: (a) the desired "smiley face" pattern, quantized to 256 gray levels. (b) The actual resistance values measured after tuning all devices with the nominal 5% accuracy, using the automated tuning algorithm[48], and (c) the corresponding statistics of the tuning errors, which is defined as normalized absolute difference between the target and actual conductance values. On panel (a), the white / black pixels correspond to 84 KΩ / 7 KΩ, measured at 0.2 V bias. The tuning was performed with 500-μs-long voltage pulses with amplitudes in a [0.8 V, 1.5 V] / [-1.8 V, -0.8 V] range to increase / decrease device conductance.

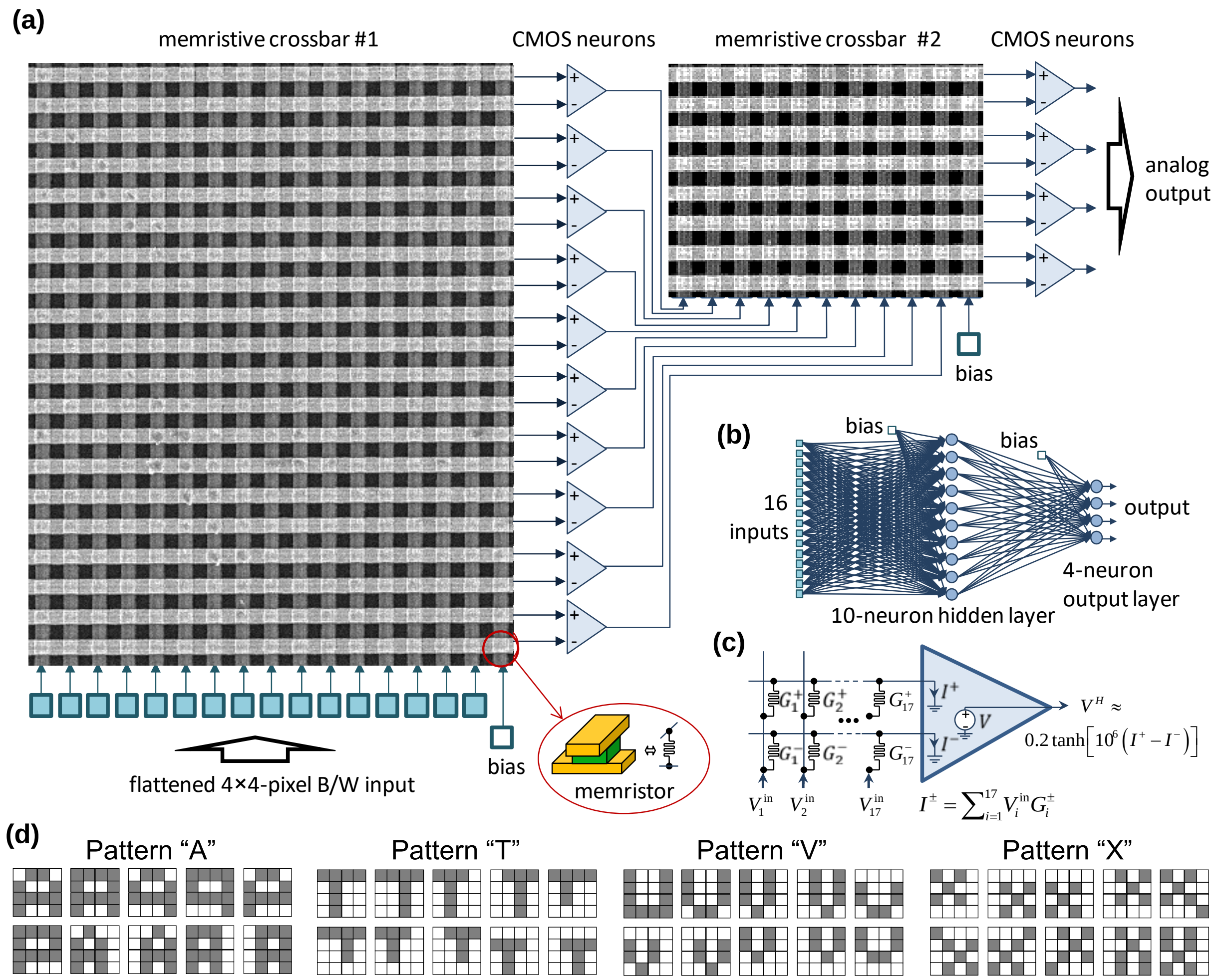

**Figure 4.** Multilayer perceptron classifier: (a) A perceptron diagram showing portions of the crossbar circuits involved in the experiment. (b) Graph representation of the implemented network; (c) Equivalent circuit for the first layer of the perceptron. For clarity only one hidden layer neuron is shown; (d) A complete set of training patterns for the 4-class experiment, stylistically representing letters "A", "T", "V" and "X".

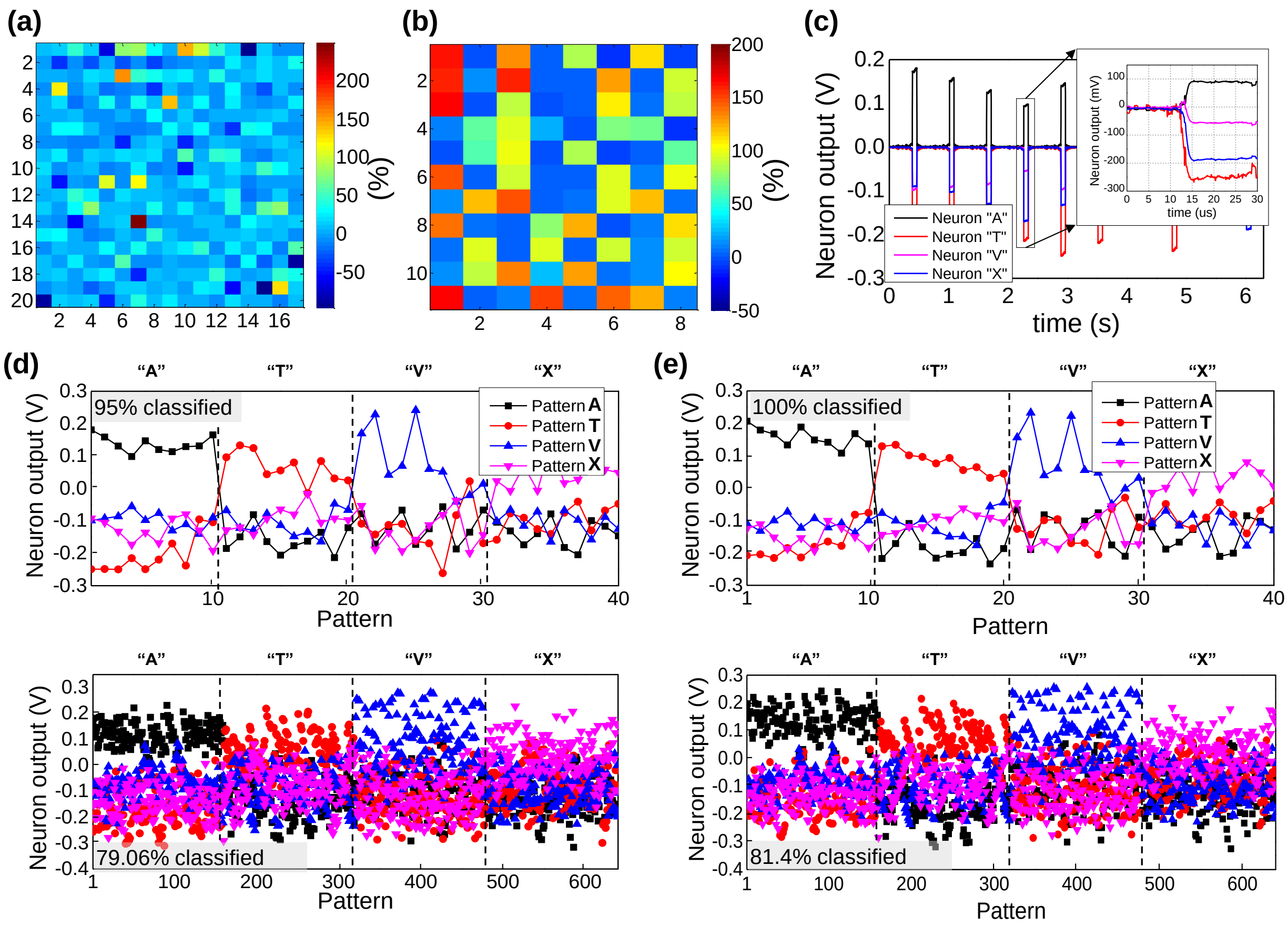

**Figure 5.** Experimental results: (a, b) The normalized difference between the target and the actual conductances after tuning in (a) the first and (b) the second layer of the network for the hardware-oblivious training approach; (c) Time response of the trained network for 6 different input patterns, in particular showing less than 5 μs propagation delay. Perceptron output voltage for (d) hardware-oblivious and (e) hardware-aware ex-situ training approaches, with top / bottom panels showing measured results for training / test patterns.

# Supplementary Information

## 1. Additional Details for the Experimental Setup

To speed up the memristor forming, a setup for its automation was developed (Fig. S1a). In general, the algorithm follows a typical manual process of applying an increasing amplitude current (or voltage) sweep to form a memristor. We have used both increasing amplitude current and voltage sweeps for forming but did not see much difference in the results of the forming procedure (Fig. 2). This could be explained by the dominant role of capacitive discharge from the crossbar line during forming, which cannot be controlled well by external current source or current compliance. (Note that to avoid overheating during voltage controlled forming, the maximum current was limited by the current compliance implemented with external transistor connected in series with biased electrode.)

Specifically, in the first step of the algorithm, the user specifies a list of crossbar devices to be formed, the number of attempts, and the algorithm parameters specific to the device technology, including the initial ($I_{start}$) and the final minimum ($I_{min}$) and maximum ($I_{max}$) values, and step size ($I_{step}$) for the current sweep, the minimum current ratio ($\mathscr{R}_{min}$), measured at 0.1 V, which user requires to register successful forming, reset voltage $V_{reset}$, and the threshold resistance of pristine devices ($R_{TH}$), measured at 0.1 V. The specified devices are then formed, one at a time, by first checking the pristine state of the device. If the measured resistance of as-fabricated memristor is lower than the defined threshold value, then the device is already effectively pre-formed by annealing. In this case, the forming procedure is not required, and the device is switched into the low conducting state to reduce leakage currents in the crossbar during the forming of the subsequent devices from the list.

Alternatively, a current sweep is applied to the device to form the device. If forming is failed, the amplitude of the maximum current in a sweep is increased and the process is repeated. (The adjustment of the maximum sweep current is performed manually for now but could be easily automated as well.) If the device could not be formed within allowed number of attempts, the same forming procedure is performed again after resetting all devices in the crossbar to the low conductive states. The second try could still result in successful forming, if the failure to form in the first was because of large leakages via on-state memristors that were already formed. Even though all formed devices are reset immediately after forming, some of them may be accidentally turned on during forming of other devices. Finally, if the device could not be formed within allowed number of attempts for the second time, it is recorded as defective.

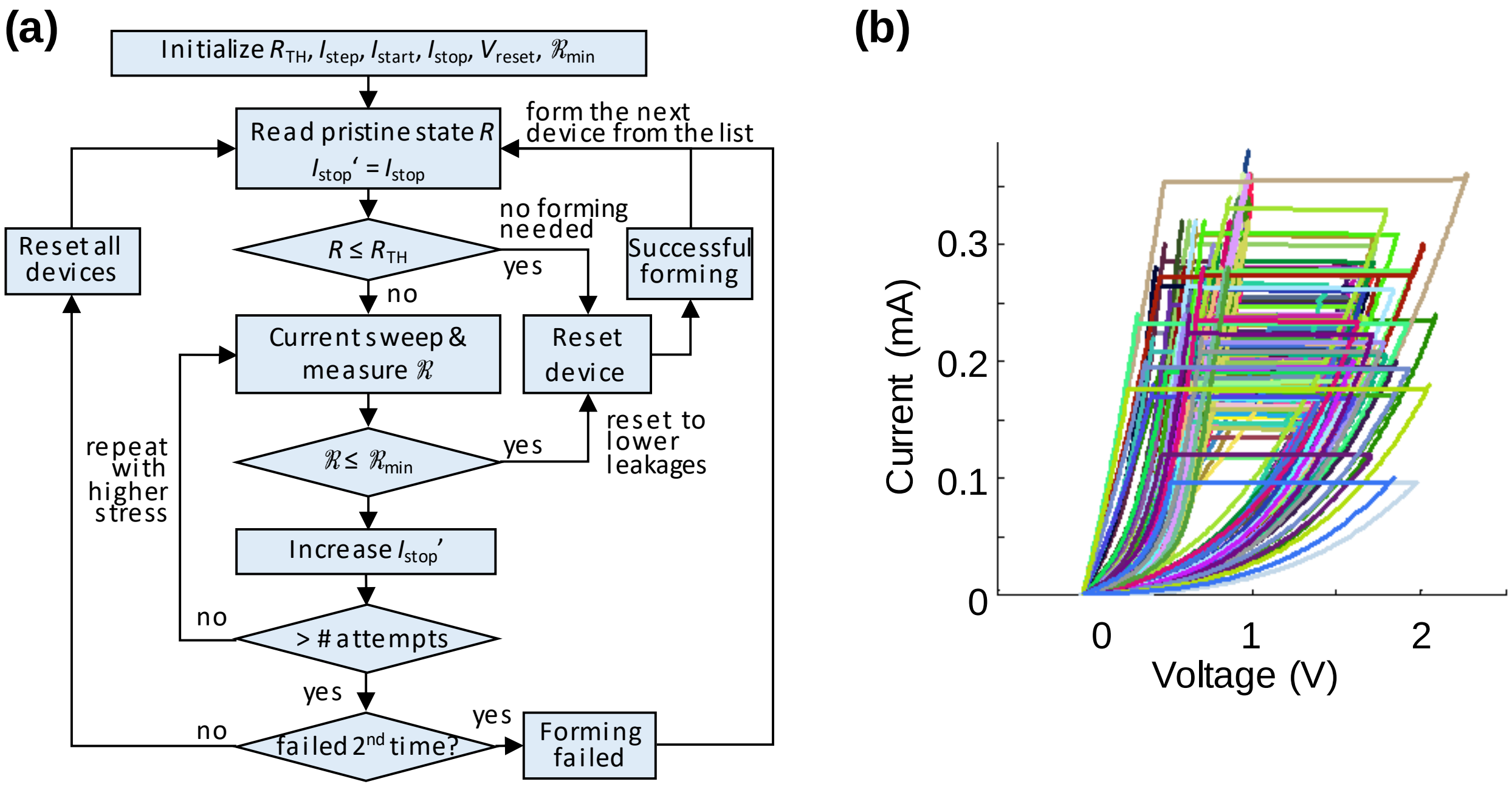

**Figure S1**. (a) Flow diagram of the automatic current-controlled memristor forming procedure. The the adjustment of $I_{stop}$’ value was so far performed manually after the failure to form a device automatically (in ~10% of all cases). (b) All forming *I*-*V* curves for one of the crossbars used in the experimental demonstration (with $I_{start}$ = 180 µA, $I_{stop}$ = 540 µA, $I_{step}$ = 20 µA, $V_{reset}$ = -1.3 V, $\mathcal{R}_{min}$ = 5).

About 1% to 2.5% of the devices in the crossbar array (i.e. 10 or less out of 400 total) could not be formed with the forming algorithm parameters that we used. It might have been possible to form even these devices by applying larger stress but we have not tried it in this experiment to avoid permanently damaging the crossbar circuit. Typically, the failed devices were stuck at some conductance state, comparable to the range of conductances utilized in the experiment, so failed device had negligible effect on the tuning accuracy.

Figure S2 shows additional details of the MLP implementation and the measurement setup. We have used AD8034 discrete operational amplifiers for the CMOS-based neurons and ADG1438 discrete analog multiplexers to implement on-board switch matrix.

Figure S3 shows absolute values of resistances and absolute error for the data presented in Figure 3.

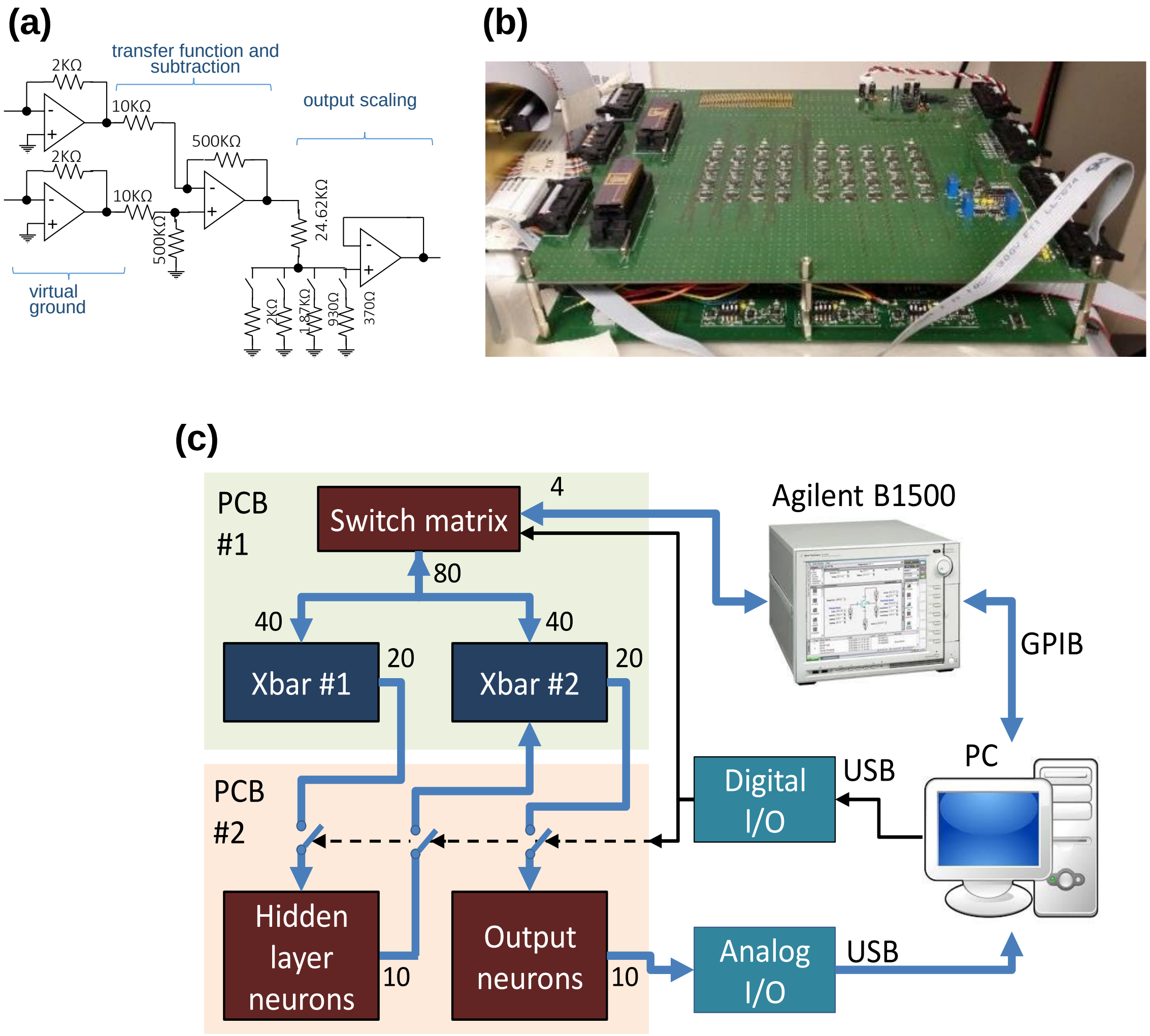

**Figure S2.** Experimental setup and board details: (a) Circuit diagram of the implemented neurons. Note that the output scaling stage is not implemented in the output neurons; (b) Photos of the two printed circuit boards with one hosting wire-bonded memristive crossbar chips and the switching matrix and the other one implementing discrete CMOS neurons; (c) Block diagram of the experimental setup controlled by a personal computer.

## 2. Additional Details and Results for Pattern Classification Experiment

Because of limited size of the classifier, we have used custom 4-class benchmark, which is comprised of a total of 40 training (Fig. 4d) and 640 test (Fig. S4) 4×4-pixel black and white patterns representing stylized letters "A", "T", "V", and "X". As Figure S5 shows, the classes of the patterns in the benchmark are not linearly separable and the use of multi-bit (analog) weights significantly improve performance for the implemented training algorithm. (The training in the software was always performed assuming neurons with hyperbolic tangent transfer function and batch-mode backpropagation training with mean-square error cost function.)

| | | | | | | | | | | | | | | | | | | | |
|---|---|---|---|---|---|---|---|---|---|---|---|---|---|---|---|---|---|---|---|
| 5.25E+04 | 8.57E+04 | 8.64E+04 | 6.94E+04 | 5.98E+04 | 6.67E+04 | 7.69E+04 | 6.27E+04 | 6.41E+04 | 5.38E+04 | 6.24E+04 | 5.09E+04 | 6.16E+04 | 6.65E+04 | 8.61E+04 | 7.38E+04 | 1.17E+05 | 9.41E+04 | 8.83E+04 | 8.63E+04 |
| 8.11E+04 | 8.82E+04 | 6.74E+04 | 5.61E+04 | 6.98E+04 | 7.46E+04 | 7.93E+04 | 7.72E+04 | 7.73E+04 | 8.86E+04 | 6.80E+04 | 8.00E+04 | 7.43E+04 | 5.64E+04 | 6.98E+04 | 8.82E+04 | 8.45E+04 | 8.23E+04 | 9.70E+04 | 9.80E+04 |
| 8.49E+04 | 6.98E+04 | 5.67E+04 | 7.81E+04 | 8.27E+04 | 5.99E+04 | 8.53E+04 | 8.80E+04 | 9.02E+04 | 9.18E+04 | 9.08E+04 | 4.01E+04 | 1.02E+05 | 6.96E+04 | 5.93E+04 | 7.15E+04 | 1.04E+05 | 8.01E+04 | 1.00E+05 | 9.85E+04 |
| 6.64E+04 | 5.91E+04 | 4.94E+04 | 8.78E+04 | 7.73E+04 | 5.11E+04 | 6.15E+04 | 7.21E+04 | 8.69E+04 | 8.51E+04 | 5.88E+04 | 5.06E+04 | 8.01E+04 | 8.15E+04 | 6.56E+04 | 5.91E+04 | 6.37E+04 | 9.41E+04 | 9.23E+04 | 8.36E+04 |
| 6.12E+04 | 6.36E+04 | 7.92E+04 | 7.54E+04 | 6.19E+04 | 1.51E+04 | 2.37E+04 | 5.29E+04 | 8.68E+04 | 5.93E+04 | 2.41E+04 | 1.72E+04 | 5.11E+04 | 8.79E+04 | 7.48E+04 | 5.67E+04 | 5.74E+04 | 9.19E+04 | 9.18E+04 | 9.22E+04 |
| 6.11E+04 | 7.11E+04 | 7.72E+04 | 7.59E+04 | 6.45E+04 | 1.49E+04 | 1.66E+05 | 5.74E+04 | 7.23E+04 | 5.78E+04 | 1.85E+04 | 8.50E+04 | 6.47E+04 | 7.15E+04 | 7.46E+04 | 6.65E+04 | 7.36E+04 | 9.08E+04 | 1.20E+05 | 9.02E+04 |
| 5.20E+04 | 6.65E+04 | 6.48E+04 | 7.73E+04 | 4.93E+04 | 1.49E+04 | 1.68E+04 | 5.16E+04 | 8.26E+04 | 5.25E+04 | 1.68E+04 | 1.65E+04 | 5.23E+04 | 7.45E+04 | 7.16E+04 | 6.69E+04 | 7.15E+04 | 6.22E+04 | 8.80E+04 | 1.03E+05 |
| 6.23E+04 | 6.46E+04 | 6.52E+04 | 1.01E+05 | 5.82E+04 | 2.46E+04 | 3.27E+04 | 5.65E+04 | 9.71E+04 | 6.47E+04 | 3.75E+04 | 2.39E+04 | 5.46E+04 | 5.46E+04 | 6.46E+04 | 6.94E+04 | 1.04E+04 | 6.52E+04 | 6.78E+04 | 9.76E+04 |
| 6.12E+04 | 6.74E+04 | 6.96E+04 | 7.05E+04 | 1.15E+05 | 5.51E+04 | 6.09E+04 | 7.04E+04 | 7.18E+04 | 7.06E+04 | 5.81E+04 | 6.48E+04 | 6.91E+04 | 6.80E+04 | 7.12E+04 | 8.19E+04 | 6.11E+04 | 6.37E+04 | 8.58E+04 | 9.45E+04 |
| 5.88E+04 | 5.73E+04 | 4.21E+04 | 3.67E+04 | 4.23E+04 | 5.20E+04 | 6.27E+04 | 5.93E+04 | 4.35E+04 | 6.03E+04 | 6.66E+04 | 5.07E+04 | 4.29E+04 | 3.39E+04 | 4.98E+04 | 6.12E+04 | 7.29E+04 | 5.13E+04 | 9.06E+04 | 9.24E+04 |
| 5.50E+04 | 7.23E+04 | 2.58E+04 | 6.81E+03 | 2.68E+04 | 3.20E+04 | 4.69E+04 | 4.23E+04 | 5.41E+04 | 4.40E+04 | 4.10E+04 | 4.22E+04 | 2.39E+04 | 7.26E+03 | 2.52E+04 | 6.02E+04 | 5.80E+04 | 6.77E+04 | 7.91E+04 | 1.00E+05 |
| 6.23E+04 | 6.57E+04 | 4.41E+04 | 1.55E+04 | 4.75E+04 | 5.52E+04 | 5.16E+04 | 5.76E+04 | 5.27E+04 | 5.10E+04 | 5.17E+04 | 5.41E+04 | 4.23E+04 | 1.64E+04 | 4.28E+04 | 7.65E+04 | 6.11E+04 | 6.40E+04 | 8.15E+04 | 1.01E+05 |
| 5.68E+04 | 6.07E+04 | 5.14E+04 | 1.67E+04 | 7.33E+03 | 1.47E+04 | 3.18E+04 | 4.36E+04 | 4.41E+04 | 3.29E+04 | 3.42E+04 | 1.66E+04 | 6.73E+03 | 2.49E+04 | 4.39E+04 | 6.06E+04 | 6.06E+04 | 5.41E+04 | 8.07E+04 | 9.32E+04 |
| 6.13E+04 | 4.91E+04 | 6.84E+04 | 4.31E+04 | 2.41E+04 | 1.59E+04 | 2.57E+04 | 2.53E+04 | 2.57E+04 | 2.40E+04 | 1.58E+04 | 1.52E+04 | 2.31E+04 | 4.99E+04 | 7.17E+04 | 5.31E+04 | 5.52E+04 | 8.37E+04 | 9.48E+04 | 1.04E+05 |
| 8.02E+04 | 5.25E+04 | 6.43E+04 | 6.55E+04 | 6.20E+04 | 3.63E+04 | 2.65E+04 | 2.74E+04 | 2.02E+04 | 2.62E+04 | 2.46E+04 | 4.34E+04 | 6.50E+04 | 6.27E+04 | 6.06E+04 | 5.18E+04 | 7.26E+04 | 1.01E+05 | 9.68E+04 | 9.53E+04 |
| 8.91E+04 | 9.28E+04 | 5.30E+04 | 6.31E+04 | 7.11E+04 | 6.17E+04 | 5.86E+04 | 4.75E+04 | 4.86E+04 | 5.12E+04 | 5.68E+04 | 6.37E+04 | 7.19E+04 | 5.06E+04 | 4.26E+04 | 7.45E+04 | 1.03E+05 | 1.07E+05 | 9.84E+04 | 8.63E+04 |
| 9.10E+04 | 7.95E+04 | 7.98E+04 | 5.75E+04 | 5.40E+04 | 5.81E+04 | 6.39E+04 | 6.86E+04 | 7.50E+04 | 7.59E+04 | 6.31E+04 | 6.09E+04 | 5.35E+04 | 8.79E+04 | 7.82E+04 | 9.04E+04 | 1.04E+05 | 9.93E+04 | 9.81E+04 | 1.02E+05 |
| 1.16E+05 | 9.55E+04 | 7.54E+04 | 8.55E+04 | 6.97E+04 | 6.62E+04 | 5.13E+04 | 4.93E+04 | 5.09E+04 | 4.98E+04 | 4.91E+04 | 6.05E+04 | 7.06E+04 | 8.32E+04 | 9.28E+04 | 9.09E+04 | 9.24E+04 | 9.68E+04 | 9.82E+04 | 9.29E+04 |
| 9.12E+04 | 9.31E+04 | 9.29E+04 | 9.51E+04 | 1.13E+05 | 4.21E+04 | 6.91E+04 | 5.33E+04 | 6.53E+04 | 8.10E+04 | 7.55E+04 | 8.94E+04 | 1.11E+05 | 9.26E+04 | 1.33E+05 | 8.92E+04 | 9.39E+04 | 9.37E+04 | 9.10E+04 | 9.40E+04 |
| 8.44E+04 | 9.53E+04 | 9.60E+04 | 8.73E+04 | 8.32E+04 | 8.22E+04 | 8.23E+04 | 1.71E+07 | 9.20E+04 | 5.01E+06 | 7.97E+04 | 6.19E+04 | 9.14E+04 | 9.66E+04 | 9.17E+04 | 9.28E+04 | 1.00E+05 | 9.72E+04 | 8.26E+04 | 7.31E+04 |
| 4.41E+04 | 1.09E+04 | 1.02E+04 | 6.25E+02 | 1.14E+03 | 3.27E+03 | -6.92E+03 | 1.57E+04 | 1.43E+04 | 2.46E+04 | 7.63E+03 | 1.91E+04 | -7.33E+02 | 3.48E+03 | 1.05E+04 | 2.28E+04 | -2.00E+04 | 2.46E+03 | 8.34E+03 | 1.03E+04 |
| 1.55E+04 | -6.81E+02 | 2.62E+03 | 4.82E+03 | 1.86E+02 | 3.77E+03 | 8.22E+03 | 1.03E+04 | 1.02E+04 | -1.07E+03 | 1.95E+04 | -1.64E+03 | -4.32E+03 | 4.53E+03 | 1.81E+02 | -7.16E+02 | 1.21E+04 | 1.43E+04 | -4.09E+02 | -1.41E+03 |
| 2.61E+03 | 1.72E+02 | 4.24E+03 | -8.08E+03 | 4.78E+03 | 2.76E+04 | 2.18E+03 | -4.87E+02 | -2.70E+03 | -4.30E+03 | -3.27E+03 | 4.74E+04 | -1.47E+04 | 3.62E+02 | 1.58E+03 | -1.52E+03 | -7.37E+03 | 1.65E+04 | -3.45E+03 | -1.93E+03 |
| 3.57E+03 | 1.78E+03 | 2.06E+04 | -2.55E+02 | 1.08E+03 | 7.31E+02 | -5.66E+02 | 6.29E+03 | 6.18E+02 | -6.72E+03 | 2.12E+03 | 1.16E+03 | -1.66E+03 | 5.98E+03 | 4.42E+03 | 1.82E+03 | 6.26E+03 | 2.50E+03 | 4.26E+03 | 1.30E+04 |
| -3.25E+02 | -2.71E+03 | -7.98E+02 | 2.97E+03 | -9.54E+02 | 1.01E+03 | 1.47E+03 | 7.98E+03 | 6.97E+02 | 1.55E+03 | 1.15E+03 | -1.09E+03 | 9.75E+03 | -9.55E+03 | 3.59E+03 | 4.19E+03 | 3.46E+03 | -4.45E+03 | 4.80E+03 | 4.39E+03 |
| -2.45E+02 | -1.14E+03 | 1.18E+03 | 2.49E+03 | -3.56E+03 | 1.15E+03 | -6.96E+04 | 3.49E+03 | 6.10E+03 | 3.10E+03 | -2.45E+03 | 1.16E+04 | -3.85E+03 | 6.92E+03 | 3.80E+03 | 3.52E+03 | -1.27E+04 | -2.08E+04 | -3.25E+04 | 6.43E+03 |
| 8.92E+03 | 3.48E+03 | 5.15E+03 | 1.12E+03 | 2.53E+03 | 1.18E+03 | -6.60E+02 | 1.68E+02 | -4.17E+03 | -6.57E+02 | -6.96E+02 | -3.91E+02 | -4.80E+02 | 3.88E+03 | -1.57E+03 | 3.10E+03 | -1.06E+04 | -1.31E+03 | -4.55E+02 | -6.04E+03 |
| -1.36E+03 | 5.40E+03 | 4.76E+03 | -3.11E+04 | 2.68E+03 | 5.50E+02 | 8.74E+02 | 4.44E+03 | -2.71E+04 | -3.81E+03 | -3.90E+03 | 1.28E+03 | 6.25E+03 | 1.54E+04 | 5.39E+03 | 5.63E+02 | 5.05E+04 | -4.30E+03 | 1.97E+04 | -9.76E+02 |
| -2.55E+02 | 2.58E+03 | 4.13E+02 | -5.33E+02 | -4.53E+04 | -3.34E+03 | 4.60E+00 | -3.94E+02 | -1.79E+03 | -6.47E+02 | 2.80E+03 | -3.90E+03 | 9.27E+02 | 2.05E+03 | -1.20E+03 | -1.19E+04 | -1.58E+02 | -2.84E+03 | 1.67E+03 | 2.07E+03 |
| 2.06E+03 | 3.57E+03 | 5.67E+02 | -3.06E+03 | 3.66E+02 | -1.65E+02 | -1.79E+03 | 1.58E+03 | 1.74E+04 | 5.80E+02 | -5.70E+03 | 1.11E+03 | -2.36E+02 | -2.56E+02 | 2.05E+03 | -2.75E+02 | -1.20E+04 | 5.41E+02 | -3.13E+03 | 4.23E+03 |
| 5.86E+03 | -1.14E+04 | -6.28E+02 | 1.87E+02 | -1.64E+03 | 1.63E+03 | -4.23E+03 | 4.18E+02 | -2.32E+03 | -1.34E+03 | 1.67E+03 | -8.59E+03 | 1.29E+03 | -2.57E+02 | 1.26E+01 | 7.47E+02 | 2.85E+03 | -6.81E+03 | 8.42E+03 | -3.68E+03 |
| -1.43E+03 | -4.76E+03 | -1.40E+03 | 6.16E+02 | -4.84E+03 | -3.43E+03 | 1.98E+02 | -5.79E+03 | -9.21E+02 | 7.97E+02 | 1.43E+02 | -2.29E+03 | 4.05E+02 | -3.09E+02 | -6.93E+01 | -6.50E+03 | -1.93E+02 | -3.14E+03 | 6.05E+03 | -4.14E+03 |
| 4.07E+03 | 1.75E+02 | 4.24E+02 | -5.82E+02 | -3.31E+02 | 1.37E+03 | 1.77E+03 | -9.17E+02 | -1.39E+03 | 7.29E+02 | -6.46E+02 | -4.84E+02 | 2.66E+02 | 3.10E+02 | 7.89E+03 | 2.98E+02 | 2.95E+02 | 2.43E+04 | 6.81E+03 | 3.44E+03 |
| -3.71E+02 | 2.75E+03 | 1.56E+03 | 8.70E+03 | 1.12E+03 | 2.08E+02 | -4.85E+02 | -1.40E+02 | -5.06E+02 | 1.21E+03 | 2.89E+02 | 8.73E+02 | 2.10E+03 | 1.95E+03 | -1.69E+03 | -1.33E+03 | 5.67E+03 | 3.82E+03 | 1.80E+03 | -7.07E+03 |
| -1.77E+03 | -6.81E+02 | -3.44E+03 | 4.48E+03 | -1.06E+03 | 6.40E+03 | -1.33E+03 | -2.17E+03 | 4.95E+03 | -1.03E+03 | 5.86E+02 | -7.48E+02 | -4.12E+03 | 7.34E+03 | 3.16E+02 | -2.26E+01 | 5.84E+03 | -4.04E+03 | -1.92E+02 | 1.34E+03 |
| 7.54E+03 | -1.44E+04 | -1.17E+03 | -2.24E+03 | -1.11E+03 | -8.14E+02 | 2.29E+03 | 4.35E+03 | 3.20E+03 | 5.80E+02 | 4.12E+03 | -2.84E+03 | -1.91E+03 | 1.19E+03 | 9.16E+03 | 3.92E+03 | -5.97E+03 | -1.02E+04 | -1.79E+03 | 1.03E+04 |
| 5.56E+03 | 1.71E+04 | -1.39E+03 | 3.41E+03 | -2.18E+03 | 2.79E+03 | -3.04E+03 | 1.39E+03 | -4.96E+03 | -5.95E+03 | -2.23E+03 | -1.60E+01 | -1.69E+03 | -2.70E+04 | 1.56E+02 | 6.19E+03 | -7.13E+03 | -2.68E+03 | -1.49E+03 | -5.12E+03 |
| -1.98E+04 | 1.10E+03 | 2.12E+04 | 2.03E+03 | 3.21E+02 | -5.27E+03 | 5.36E+02 | 2.46E+03 | 8.85E+02 | 1.96E+03 | 2.66E+03 | 4.14E+02 | -5.79E+02 | 4.25E+03 | 3.79E+03 | 5.69E+03 | 4.21E+03 | -1.84E+02 | -1.60E+03 | 3.72E+03 |
| 5.41E+03 | 3.49E+03 | 3.72E+03 | 1.54E+03 | -1.60E+04 | 4.54E+04 | 9.34E+03 | 1.67E+04 | 4.72E+03 | -1.10E+04 | 2.92E+03 | -1.86E+03 | -1.42E+04 | 4.02E+03 | -3.65E+04 | 7.37E+03 | 2.74E+03 | 2.94E+03 | 5.63E+03 | 2.60E+03 |
| 1.22E+04 | 1.34E+03 | 6.22E+02 | 9.32E+03 | 1.34E+04 | 5.29E+03 | 5.21E+03 | -1.70E+07 | -4.49E+03 | -4.92E+06 | 7.80E+03 | 2.56E+04 | 5.24E+03 | -5.00E-01 | 4.94E+03 | 3.84E+03 | -3.79E+03 | -5.92E+02 | 1.40E+04 | 2.35E+04 |

**Figure S3**: Absolute device resistances (top) and absolute tuning error (bottom) for the smiley face experiment.

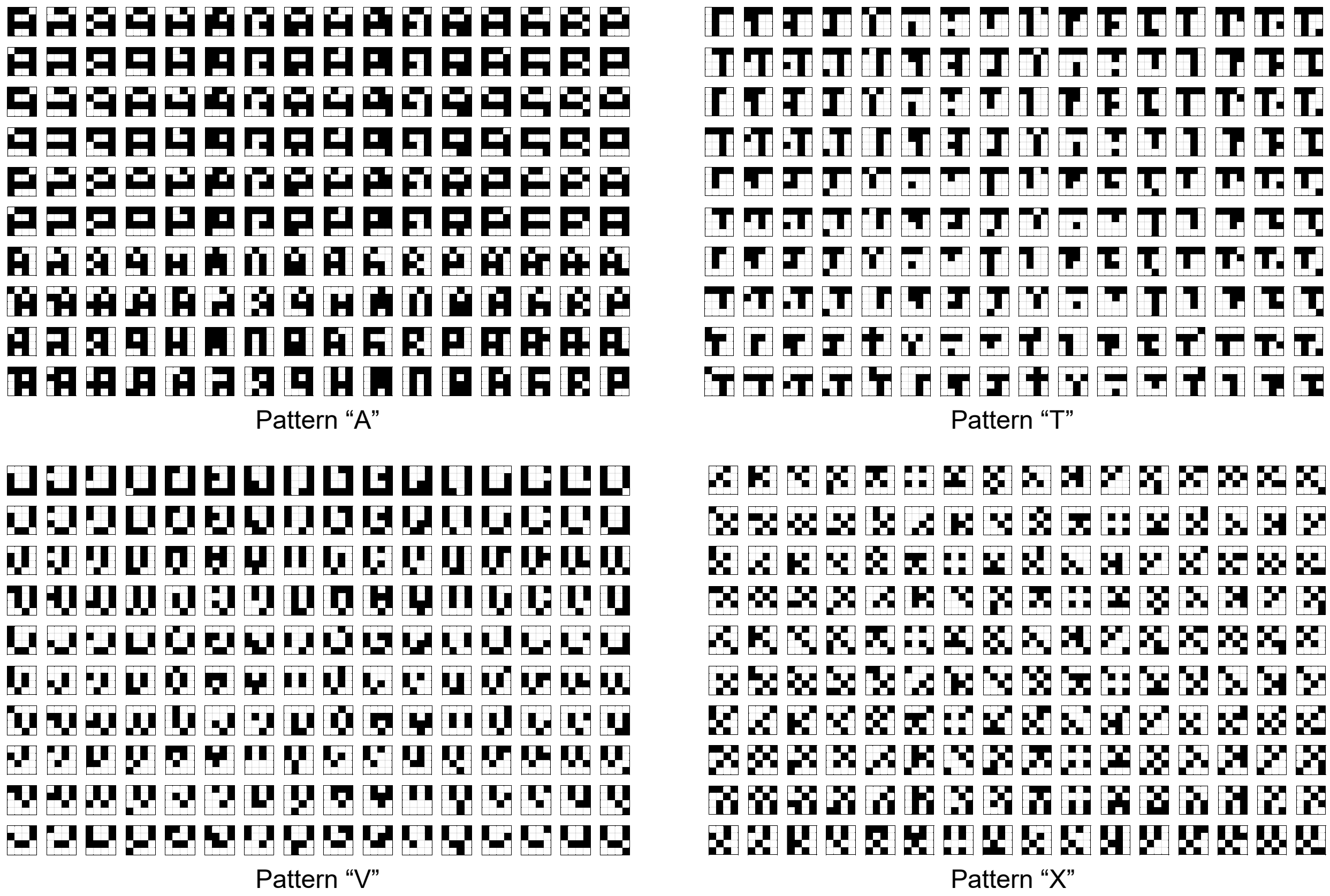

**Figure S4.** A complete set of 640 test patterns.

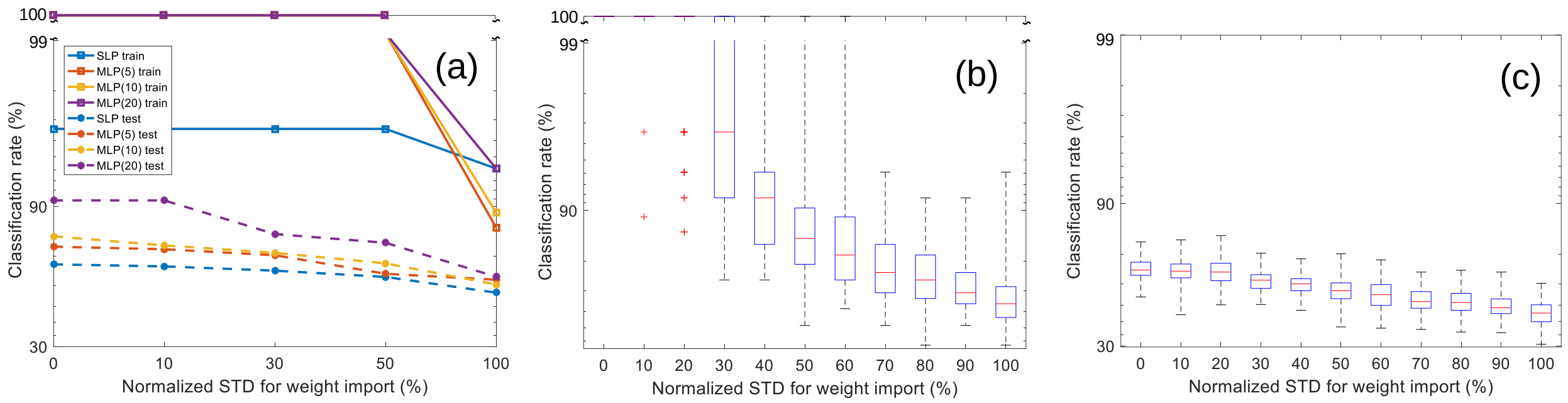

**Figure S5.** Simulated classification performance for several networks as function of weight import precisions: (a) Comparison of the best fidelity obtained for single layer perceptron and MLPs with different number of hidden layer neurons (shown in parenthesis in the legend). (b, c) The results for 10-hidden layer perceptron, similar to the one used in the experiment for classification of (b) training and (c) test patterns. The weight error was modeled by adding, to its optimized value, a normally distributed noise with the shown standard deviation. The red, blue (rectangles), and black (segment) markers denote, respectively, the median, the 25%-75% percentile, and the minimum and maximum values for 100 simulation runs.

In addition to ex-situ method (see Figure 5 of the main text and its discussion), we have also trained the network "in-situ", i.e. directly in a hardware [1, 2]. (Similar to our previous work [1, 2], only inference stage was performed in a hardware during such in-situ training, while other operations, such as computing and storing the necessary weight updates, were assisted by an external computer.) Unfortunately, because of limitations of our current experimental setup, we could only implement in-situ training using fixed-amplitude training pulses (which is similar to Manhattan rule). The classification performance for this method was always worse as compared to that of both hardware-aware and hardware-oblivious ex-situ approaches. For example, the experimentally measured fidelity for 3-pattern classification task was 70% (Fig. S6), as compared to 100% classification performance achieved on training set using both ex-situ approaches. This is expected because in ex-situ training the feedback from read measurements of the tuning algorithm allows to effectively cope with switching threshold variations by uniquely adjusting write pulse amplitude for each memristor, which is not the case for the fixed-amplitude weight update (Fig. S6). We expect that fidelity of in-situ trained network can be further improved using either variable-amplitude implementation [3].

### 3. Crossbar Circuit Scaling

An important future work, in addition to the monolithic integration with CMOS subsystem discussed in the main text, is increasing the dimensions of the crossbar circuits which would allow higher connectivity among neurons and improve integration density (i.e. by lowering relative peripheral overhead). Here let us first stress again that in our implementation, crossbar lines are never floated so that sneak path currents do not affect directly the measured currents at the outputs. Scaling up crossbar dimensions, however, increases currents flowing in the crossbar lines. Because of the potential voltage drops across the crossbar lines the voltages applied to the crosspoint memristors could be different from the ones applied at the periphery.

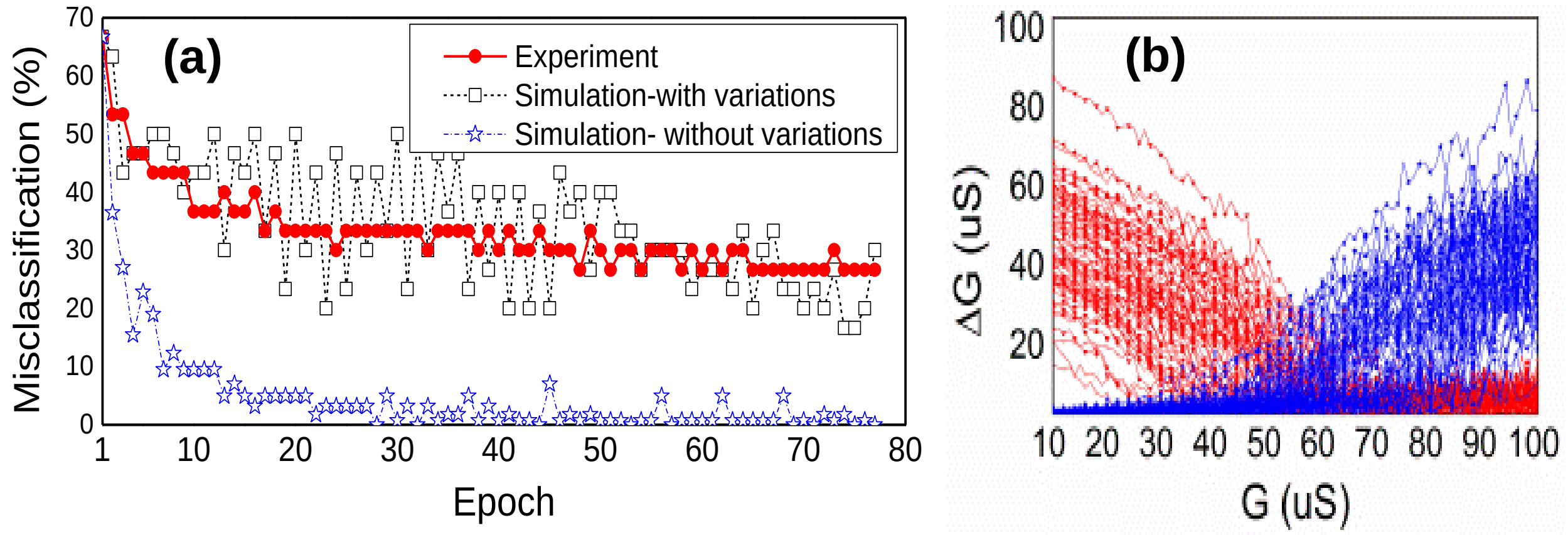

**Figure S6.** In-situ training for 3-pattern classification ("A", "V", and "T"): (a) Experimentally measured and simulated error decay dynamics for the training set patterns. In experiment, conductances of all memristors were updated, one row of the crossbar at a time, at the end of each epoch. The weight update in each row was done in parallel in two steps by applying 500-μs fixed amplitude (± 1.3 V) voltage pulses using *V*/2 biasing technique. (b) Example of devices' switching kinetics and it's variations obtained using simple device model from Ref. [1]. Such model was used for the in-situ training simulations shown in panel a – see supplementary matlab code for more details.

For example, Figure S7 shows the dependence of the worse-case voltage drop as a function of the length of the finite resistor ladder, which is useful for analyzing crossbar circuit operation. In this figure, one set of lines shows the voltage drop assuming electrode resistance per wire segment ($R_w$) comparable to the one in our experiment, while the other one is for more aggressive (though quite realistic) parameters which are representative of high-aspect ratio copper wires. For simplicity, the memristor conductances $G(V)$ can be estimated using the corresponding average value measured at bias $V$, specific to the type of considered operation. It should be noted that in a properly trained network, the weights are typically normally distributed so that the representative average value is rather close to the minimum of the used conductance range.

Let us now consider in detail three operations which might be impacted by voltage drop on the crossbar lines, namely classifier inference, and read and write phases of the tuning algorithm:

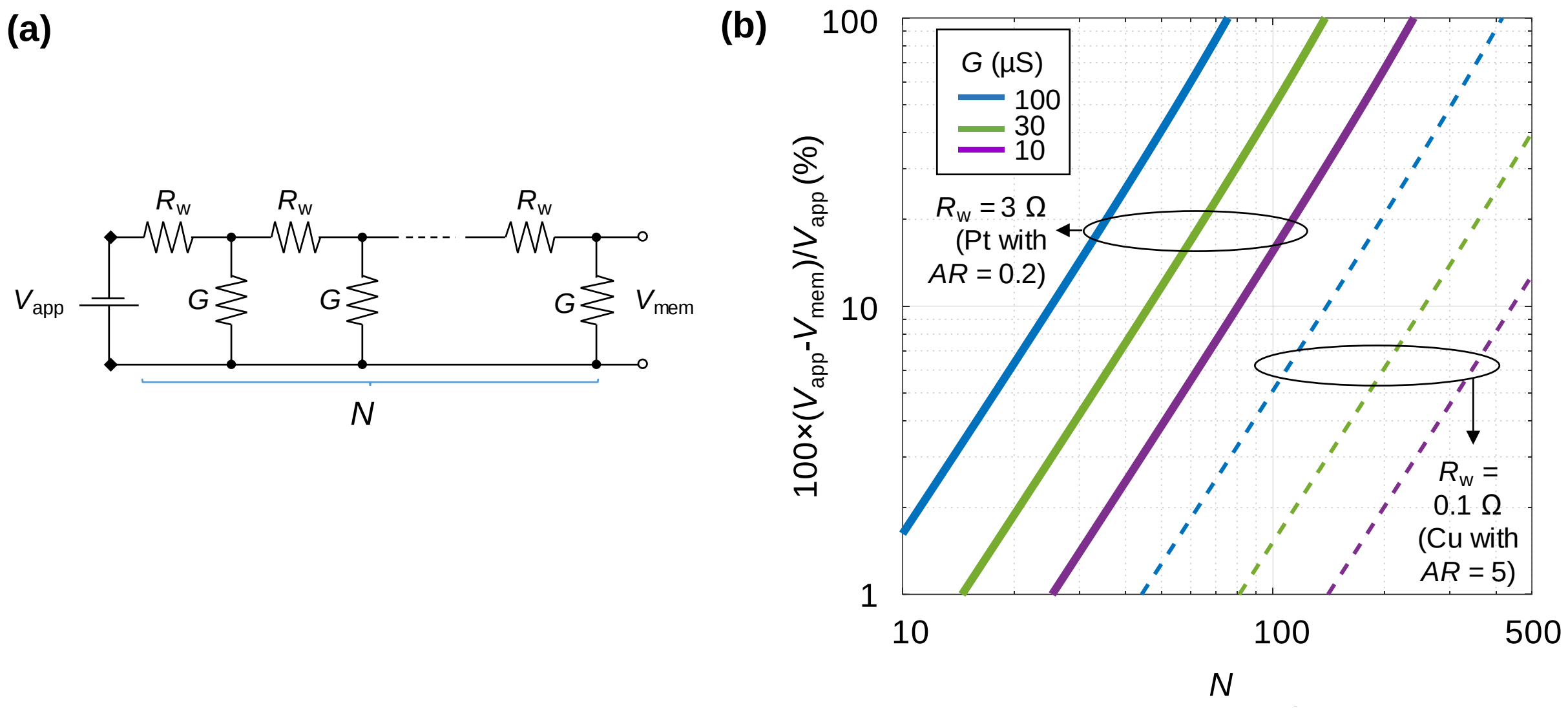

**Figure S7.** Voltage drop in resistor ladder: (a) The considered circuit and (b) the relative worst-case voltage drop for several representative parameters specific to the implemented crossbar circuits. *AR* stands for the electrode height-to-width aspect ratio.

- *Write operation*

Naturally, the voltage drops are the most significant for write operation because of the larger voltages applied and higher currents passed. For the conductance tuning, however, we do not rely on precise conductance update with write pulses but rather adjust applied write voltages gradually based on precise read measurements. Therefore, any potential voltage drop will be compensated dynamically during tuning by applying larger voltage pulses, with the largest applied voltage (and hence crossbar dimensions) limited by the condition of not disturbing half-selected devices.

Specifically, let us assume the *V*/3 biasing scheme, i.e. with $\pm V_W/2$ applied to the selected lines and $\pm V_W/6$ to the remaining lines. From Fig. 1c and 2, up to $(V_{TH}^{SET})_{max} \approx +1.3$ V set and $(V_{TH}^{RESET})_{max} \approx -1.9$ V reset voltages must be applied to switch the devices with the largest switching thresholds. (Here, we neglect the tails of the distributions on Fig. 2, which are typically contributed by the devices at the edges of the array. This is similar to the dummy line technique commonly used in conventional memories.) The corresponding average memristor conductances at one third of such biases can be roughly estimated to be $\langle G((V_{TH}^{SET})_{max}/3)\rangle \approx 30$ μS for set and $\langle G((V_{TH}^{SET})_{max}/3)\rangle \approx 50$ μS for reset transitions. On the other hand, the largest voltages, which can be safely applied to the half-selected devices without disturbing memristors with the smallest switching thresholds are $(V_{TH}^{SET})_{min} \approx +0.7$ V for set and $(V_{TH}^{RESET})_{min} \approx$ -1 V for reset transitions. The maximum crossbar dimensions, specific to the wire resistance, memristor *I*-*V* and its variations (i.e. parameters $R_w$, $G((V_{TH})_{max}/3)$, $(V_{TH})_{max/min}$ ) can be crudely estimated assuming $100\times(3(V_{TH})_{min} - (V_{TH})_{max})/(V_{TH})_{max} / 2$ as the largest allowable relative voltage drop in Fig. S7b. (Additional factor of 2 in the denominator accounts for the drop on both selected lines.) For the considered parameters, this drop is equal to 30% and 25% for set and reset switching, respectively, indicating to the possibility of implementing 70×70 crossbar arrays with demonstrated device technology and up to 400×400 crossbar array for the crossbar arrays with improved electrode resistance. (Note that in our work, we have used somewhat simpler, the *V*/2 biasing scheme, for which the largest allowable voltage drop is ~ 7% and the corresponding maximum crossbar dimensions are around 40×40 and 200×200 for two considered electrode resistances.)

- *Read operation*

Let us assume that during read operation, one of the selected lines is biased at $+V_R$, while the other selected line and all of the remaining ones are grounded. (This is exactly the scheme that we used for conductance tuning in this work.) In this case, the current running via grounded selected crossbar line is small (only contributed by one selected memristor) and does not dependent on the crossbar dimensions. Therefore, the substantial voltage drops may occur only on the biased selected crossbar line. Such voltage drop would be naturally much less than that of the write operation and, moreover, it can be easily taken into account when reading the state of the devices. For example, it is straightforward to compute the actual applied voltage across the specific memristor knowing the conductive states of all other half-selected devices of the biased selected crossbar electrode.

- *Inference operation*

As discussed in main text, during inference, one set of lines (vertical in Figure 3a) receive voltages $V \leq V_R$, while all orthogonal lines are virtually grounded. Because of the smaller applied voltages, the crossbar line currents, and hence the corresponding voltage drops, are the smallest for inference operations. However, the inference operation (just like read) is more sensitive as compared to write operation to the voltage variations and even small voltage drops may lead to the lower effective precision of the vector-by-matrix computation. For example, assuming representative 10 μS average device conductance, and 70×70 and 400×400 crossbar arrays discussed in write operation above, the worst-case voltage drop on one line is around 7% (Fig. S7b).

Using our examples, inference operations would likely be a limiting factor for scaling though are several reserves for improvements. For example, the conductances of each memristor can be uniquely increased to compensate for the potential voltage drops during inference. (Unlike read operation, such adjustment cannot be exact because of the input-dependent voltage drop on the virtually-grounded lines.) The loss of precision for the worst case largest currents might be also acceptable, e.g. if it leads to the saturation of the neuron. It is also important to note that precision loss at inference due to voltage drops is common problem for the devices with or without selectors. If fact, the problem is likely more severe for 1T1R structures, because of their larger device area and potentially larger $R_w$.

The crude estimate above show that the developed device technology, with some further optimization of the electrodes, should be suitable for implementing much larger, up to 400×400 crossbar circuit. The discussed analysis is also applicable to 10 nm memristors, if we assume that both the resistance of the crossbar line segment and memristor operating (average) currents would scale down at the same rate. (For that memristor currents should decrease at slightly faster rate than its linear device dimensions to compensate for the additional increase in metal resistivity due to scattering effects.) That is certainly plausible scenario for smaller currents at voltages below $V_R$ (e.g., relevant to the inference operation and read phase of the tuning algorithm) considering that the off-state conductance is typically limited by the device leakages which are proportional to the device electrode area. Ensuring the same scaling in the context of the write phase of the tuning algorithm would require enhancing *I-V* nonlinearity and/or decreasing write currents, which we believe is also plausible given the observed write current dependence on the electrode area in our devices and further optimization of the tunneling barrier layer.

## 4. Temperature Sensitivity

Practical neuromorphic hardware should be able to operate correctly under wide temperature ranges. Even though we have not measured experimentally the sensitivity of the functional performance to the ambient temperature variations let us note that for the proposed circuits, the change in memristor conductance with ambient temperature (Fig. S8a, b) is already partially compensated because of differential synapses implementation, with each weight $W \equiv G^{+} - G^{-}$ implemented with a pair of memristors with conductances $G^{\pm}(T) = G_{\mathrm{BIAS}} \pm G/2$. Noting that temperature dependence is the weakest for higher conductive states (Fig. S8a, b), the temperature dependence can be further reduced by implementing weights with higher conductances, i.e. choosing larger values of $G_{\mathrm{BIAS}}$.

Additional approach to reduce sensitivity is to utilize memristor (with conductance $G_{\mathrm{M}}$) in the feedback of the second opamp stage of the original neuron circuit (Fig. S2a). In this case, the output of the second stage is proportional to $\Sigma_i V_i^{\mathrm{in}}(G_i^{+}\text{-}G_i^{-})/G_{\mathrm{M}}$ with temperate drift additionally compensated assuming similar temperature dependence for the feedback memristor.

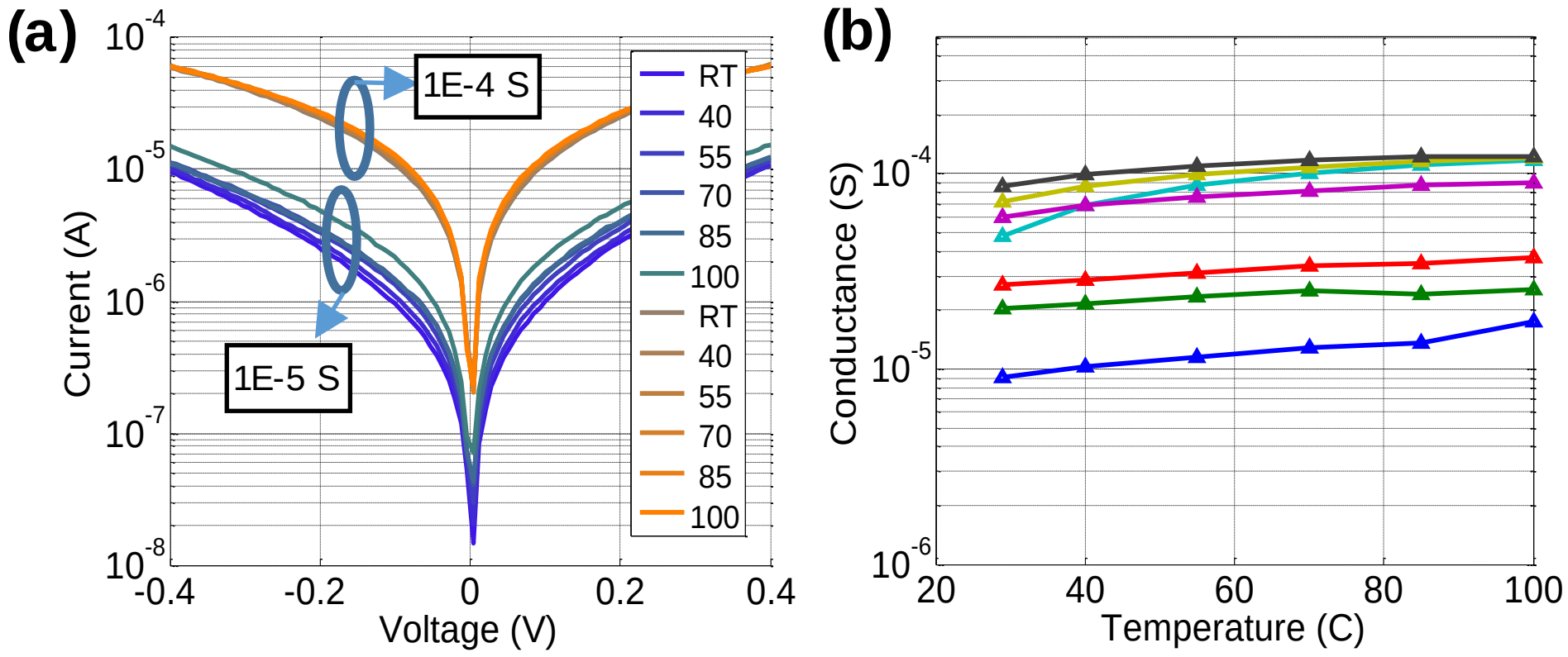

**Figure S8.** Preliminary temperature sensitivity study: (a) The *I-V* curves of a single memristor for several temperatures and (b) the extracted temperature dependence of its conductance.